\documentclass[]{raa}            % referee version: for submission

\usepackage{graphicx,times}             %for PS/EPS graphics inclusion, new
\usepackage{color}
\usepackage{soul}

\begin{document}

   \title{Period variations of Algol-type eclipsing binaries AD~And, TW Cas and IV~Cas
%\,$^*$
%\footnotetext{$*$ Supported by the National Natural Science Foundation of China.}
}
%   \subtitle{I. Place Your Subtitle Here}

   \volnopage{Vol.0 (20xx) No.0, 000--000}      %%preserved for Editor. DOn't remove!
   \setcounter{page}{1}          %%starting page, preserved for Editor. DOn't remove!

   \author{\v{S}. Parimucha
      \inst{1}
   \and P. Gajdo\v{s}
      \inst{1}
   \and V. Kudak
      \inst{1}
   \and M. Fedurco
      \inst{1}      
   \and M. Va\v{n}ko
      \inst{2}
   }
%% Here is an example of three authors come from different institutes.
%% For single author or all the authors from an institute, use "\inst{}" only

   \institute{Institute of Physics, Faculty of Science, University of P. J. \v{S}af\'{a}rik, 044~01 Ko\v{s}ice, 
	      Slovakia; {\it stefan.parimucha@upjs.sk}\\
   \and
	      Astronomical Institute, Slovak Academy of Sciences, 059~60 Tatransk\'a
	      Lomnica, Slovakia\\        
   }

   \date{Received~~2017 month day; accepted~~2017~~month day}

\abstract{ We present new analysis of $O-C$ diagrams variations of three Algol-type eclipsing binary stars  AD~And, TW~Cas and IV~Cas. We have used all published minima times (including visual and photographic) as well as new determined ones from our and SuperWasp observations. We determined orbital parameters of the 3$^{rd}$ bodies in the systems with statistically significant errors,  using our code based on genetic algorithms and Markov Chain Monte Carlo simulation. We confirmed multiple nature of AD~And and triple-star model of TW~Cas and we proposed quadruple-star model of IV~Cas. 
\keywords{binaries:close -- binaries: eclipsing -- techniques: photometric -- stars: individual: AD~And, TW~Cas and IV~Cas}
}

   \authorrunning{Parimucha et al.}            %author_head in even pages
   \titlerunning{Period variations of AD~And, TW Cas and IV~Cas }  % title_head in odd pages

   \maketitle
%% The author head (on even pages) and the title head (on odd pages) will be
%% automatically extracted from \author{} and \title{}. Whenever the title is too long,
%% you will be asked to supply a shorter one by inserting either \authorrunning{} or
%% \titlerunning{} before \maketitle. Anyway, you can specify your own heads.
%%
%%
%% Note: In the following text body of your manuscript, please note several differences from
%%       other major journals:
%% (1) \subsection{Please Capitalize the First Letter of Each Notional Word in Subsection Title}
%% (2) Please Capitalize the First Letter of Each Notional Word in all tables' captions

%
%________________________________________________ sections below
%
\section{Introduction}           %% first-level sections will be auto-capitalized
\label{sec:introduction}

Study of $O-C$  diagrams of eclipsing binaries is a powerful tool for an analysis of  temporal variations and irregularities in the cyclic phenomena observed in the stars. The most frequently determined quantity in the period studies are minima times of the binary light curve. If we determine minimum time ($O$ -- observed), we can calculate difference between this value and predicted by the ephemeris ($C$ -- calculated). If the changes of $O-C$ values with time are systematic and if they exceed the experimental errors we can provide a better model of a such system and reveal other hidden physical phenomena, like a mass transfer between the both components, angular momentum lost from the system, apsidal motion  and/or presence of another body in the system (Sterken~\cite{sterken2005}).

In this paper we present a new period analysis of 3 Algol-type eclipsing binaries, which have been overlooked for few past
years. 

The light variability of {\bf AD And} was discovered by Gutnik \& Preger~(\cite{gutnik1927}). They classified the variations as $\beta$~Lyr type with a photographic amplitude about 0.9~mag. The first photometric study of AD And was published by Taylor \& Alexander~(\cite{taylor1940}). Rucinski~(\cite{rucinsky1966}) published the first photoelectric photometry of the object and determined 5 minima times. 
Cannon~(\cite{cannon1934}) classified star as a F-type object, later classification of Hill et al.~(\cite{hill1975}) gave spectral type range from B8 to A0. Giuricin \& Mardirossian~(\cite{giuricin1981}) published photometric parameters of the system and concluded that the both components have  almost the same radii, masses, temperatures and luminosities with orbital inclination $i=81.9\pm0.4$\degr, what was confirmed by Liakos et al.~(\cite{liakos2012}). The period variations of AD And were investigated by several authors (Whitney~\cite{whitney1957}; Rucinsky~\cite{rucinsky1966}; Frieboes-Conde \& Herczeg ~\cite{Frieboes1973};  Liao \& Qian~\cite{liao2009}; Liakos et al.~\cite{liakos2012}). The last mentioned authors determined period of the third body to be 14.3 years and its mass function $f(m)=0.183$~M$_{\odot}$.

\begin{table*}
\caption{Basic parameters of studied objects. TYC - number in Tycho-2 catalog, $V$ - magnitude in the V filter, $B-V$ - colour index, $T_0$ - initial time of minimum, $P$ - period. Linear ephemeris are taken from the on-line database  (Kreiner~\cite{kreiner2004}).}
\label{tab:parameters}
\begin{center}
 
 \begin{tabular}{lccccc}
  \hline\noalign{\smallskip}
    Star     	&  TYC  	& $V$[mag] 	& $B-V$[mag] & $T_0$ [HJD] 	& $P$ [d]   \\
  \hline\noalign{\smallskip}
  AD And  	& 3641-0151-1  	& 11.14 & 0.20  & 2452500.3670  & 0.9862210 \\ 
  TW Cas  	& 4059-0898-1  	& ~8.32 & 0.10  & 2452500.8245  & 1.4283346 \\  
  IV Cas  	& 4001-1104-1  	& 11.34 & 0.27  & 2452500.3569  & 0.9985067  \\
  \hline\noalign{\smallskip} 
 \end{tabular}

\end{center}
\end{table*} 

\textbf{TW Cas} was discovered in 1907 by Pickering~(\cite{pickering1907}) and its variability from photographic observations was confirmed by Zinner~(\cite{zinner1913}). Spectroscopic observations of Struve~(\cite{struve1950}) confirmed B9 spectral type of the primary component and he determined mass function $f(m)$=0.098 M$_\odot$. The most recent photoelectric $V$ observations of TW Cas were obtained by Narita et al.~(\cite{narita2001}). Their light-curve solution led to conclusion that the secondary component almost fills its Roche lobe.
Djurasevic et al.~(\cite{djurasevic2006}) re-analysed older photoelectric observations from McCook~(\cite{mccook1971}) and determined masses of the primary and secondary components to be $M_{1}$=2.66 M$_\odot$ and $M_{2}$=1.15 M$_\odot$, respectively, which is in agreement with values obtained by Narita et al.~(\cite{narita2001}). Kreiner~(\cite{krainer1971}) used all the available minima times of TW Cas, but could draw no definite conclusions concerning the period variations. Narita et al.~(\cite{narita2001}) assumed that the orbital period of TW Cas was slowly decreasing, as was confirmed by
Lloyd \& Guilbault~(\cite{lloyd2002}). Khaliullina~(\cite{kaliulina2015}) noted that the recent minima times demonstrate sinusoidal changes of the orbital period and thus, cyclic variations of the period due to the presence of a third body in the system are observed.

Eclipsing binary {\bf IV Cas} was discovered on Moscow photographic plates by Meshkova~(\cite{meshova1940}). Kim et al.~(\cite{kim2005}) in their photometric study discovered a short-periodic pulsating component with a frequency of 37.672 cycles per day (period $\sim$38 minutes). Wolf et al.~(\cite{wolf2006}) and Zasche~(\cite{zasche2006}) reported sinusoidal $O-C$ diagram changes caused by light-time effect with period about 21800 days and semi-amplitude 0.03 day. The third component should have a minimal mass of $0.96$~M$_{\odot}$. Detailed analysis of the binary light curve as well as pulsation characteristics of the primary component was studied by Kim et al.~(\cite{kim2010}). They showed that IV~Cas is in a semi-detached configuration with A3 spectral type primary component and evolved early-K secondary, which fills its inner Roche lobe. Pulsations correspond to $\delta$~Scuti-type pulsator.

The basic parameters of the studied stars, like a their brightness, colour indices and linear ephemeris are given in Tab.~\ref{tab:parameters}.

\begin{table}[t]
\begin{center}
\caption{New unpublished minima times JD$_{min}$ [HJD - 2400000] of AD~And, TW~Cas and minima times determined from SuperWASP observations of IV~Cas. Errors of minima times determinations are given in parenthesis.} \label{tab:minima}
\begin{tabular}{lcc}
  \hline\noalign{\smallskip}  
  & JD$_{min}$ & JD$_{min}$\\
  \hline\noalign{\smallskip}
 AD And  &  57966.4661(2) & 		 \\
 TW Cas  &  57948.4948(2) &   		\\  

  \hline\noalign{\smallskip}   
IV Cas  & 54319.6353(3)  & 54361.5721(2)  \\ 
	& 54337.6076(5)  & 54362.5705(2)  \\
	& 54348.5919(2)  & 54363.5691(2)  \\
	& 54350.5890(4)  & 54381.5421(3)  \\
        & 54351.5873(3)  & 54382.5405(3)  \\
	& 54352.5853(2)  & 54396.5202(3)  \\       
	& 54353.5835(3)  & 54398.5169(4)  \\
	& 54354.5828(2)  & 54399.5157(7)  \\
        & 54357.5784(3)  & 54402.5113(2)  \\ 
        & 54360.5734(3)  & 54405.5063(3)  \\   
  \hline\noalign{\smallskip}
                              
\end{tabular}                                      
\end{center}    
\end{table}

\section{Minima times}
\label{sec:minima}

For our analysis we used minima times collected into on-line database O − C gateway operated by the Czech Astronomical Society\footnote{http://var2.astro.cz/ocgate/}. Almost all published minima times of our objects (including visual and photographic) are accessible in this database.

We used also our two new unpublished minima times of AD~And and TW~Cas. Moreover, we determined minima times from SuperWasp project observations (Pollaco et al.~\cite{pollaco2006}) available from public archive\footnote{http://wasp.cerit-sc.cz/form}. Our observations were obtained using 508mm telescope operated by University (Parimucha \& Va\v{n}ko~\cite{parimucha2015}). Data reduction and differential photometry was performed by C-Munipack package\footnote{http://c-munipack.sourceforge.net/}. 

\begin{figure}[t]
 \includegraphics[width=\columnwidth]{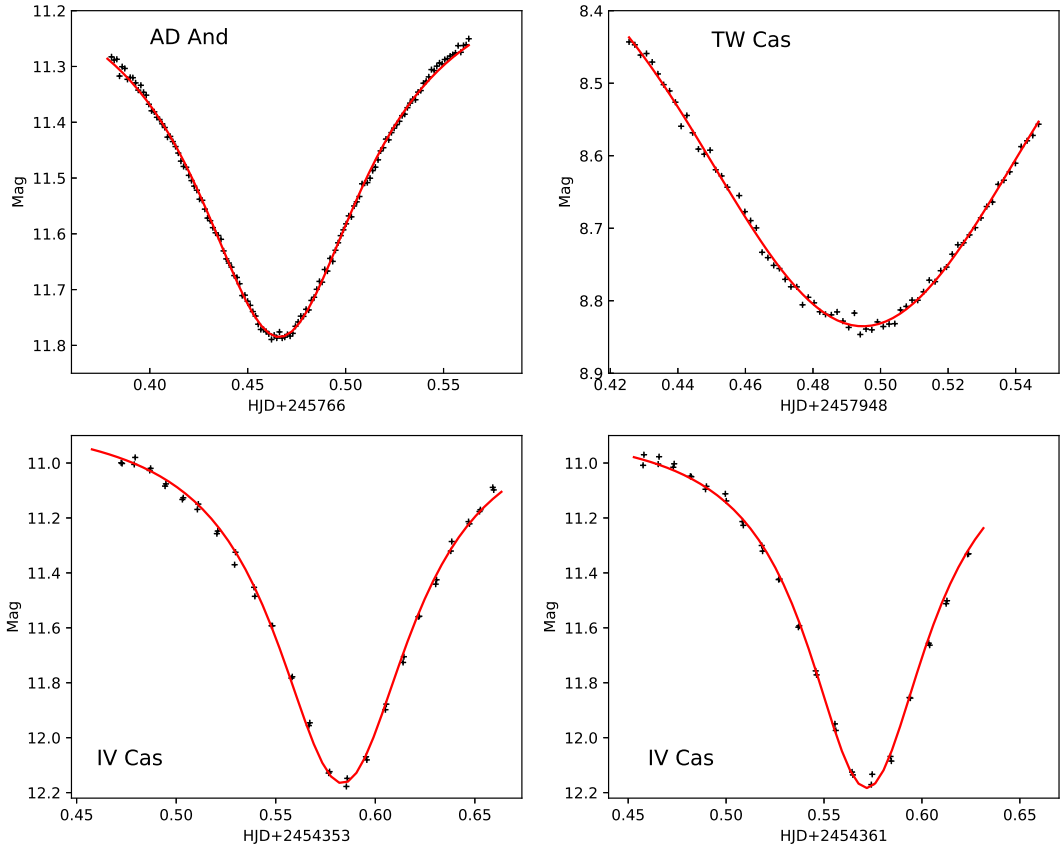}  
 \caption{Our new unpublished light curves of AD And and TW Cas from 2017 and examples of two light curves of IV~Cas from the SuperWasp archive. The best fit with template function for each minimum is displayed by the red line.}
 \label{fig:04}  
 \end{figure}

New minima times were calculated by the fitting to template function of the minimum light-curve as proposed by Mikul\'a\v{s}ek~(\cite{mikulasek2015}). These minima times are listed in Table \ref{tab:minima}. Our new light curves of AD~And and TW Cas, together with examples of two SuperWasp light curves of IV~Cas are displayed in Fig.~\ref{fig:04}. The best fit with template function for each minimum is also depicted by the red line.

\section{Data analysis}
\subsection{Theory}
Minima times $T_C$ of eclipsing binary stars can be simply calculated by the linear ephemeris:
\begin{equation}
\label{eq:linear}
 T_C = T_0+ P \times E,
\end{equation}
which predicts minima times of eclipsing binary with an orbital period $P$ without any other influences. Here $E$ is an epoch of the observation (integer number for a primary minimum and/or $E+0.5$ for a secondary minimum) and it counts, how many eclipses elapsed since the zero epoch. $T_0$ is an initial minimum time (minimum at $E=0$). Up to date linear ephemeris for our objects are listed in Table \ref{tab:parameters}. Difference between observed $T_O$ and predicted $T_C$ minima times is caused by perturbation $\delta T$
\begin{equation}
T_O - T_C \equiv O-C  = \delta T.
\end{equation}

This perturbation is generally a sum of different effects. For our analysis we consider only mass transfer and presence of the other third body in the system (light-time effect). Than we can write
\begin{equation}
\delta T =  Q \times E^2 + \frac{a\sin i_3}{c} \left[\frac{1-e_{3}^2}{1+e_3\cos\nu_3}\sin(\nu_3+\omega_3)+e_3\sin\omega_3\right].     
\label{eq:dt}
\end{equation}
The first term represents period change due to mass transfer (Hilditch~\cite{hilditch2001}). The second one describes period change due to light-time effect caused by the third component (Irwin~\cite{irwin1952}). Here is $a\sin i_3$ projected semi-major axis of the orbit with eccentricity $e_3$, $c$ is a speed of light, $\omega_3$ is the longitude of the periastron and $\nu_3$ is the true anomaly of the binary orbit around the center of the mass of the system. There are no limitations to mass or orbital parameters of the third body. Period of the third body $P_3$ and the time of pericenter passage $t_{03}$ are hidden in $\nu_3$ calculation, which have to be solved using Kepler equation. Because we are not able to find inclination of the orbit $i_3$ only from $O-C$ analysis, we can determine only so-called mass function of the third body:
\begin{equation}
\label{eq:massfun}
f(M_3) = \frac{(M_3 \sin i_3)^3}{M^2} = \frac{(a \sin i_3)^3}{P_3^2},
\end{equation}
where $M = M_1+M_2+M_3$ is a total mass of the system ($M_i$ - masses of components).
	
\subsection{Fitting method}    
To obtain the optimal set of 8 parameters ($T_0$, $P$, $Q$, $t_{03}$, $P_3$, $a\sin i_3$, $e_3$, $\omega_3$) there are in use classical numerical methods based on the iterative minimization of the sum of squares, like a Levenberg-Marquardt algorithm or Simplex method (Press et al.~\cite{press2007}). These algorithms can be simply implemented in many programming languages and data analysis packages and solution can be found relatively fast. But the convergence to the global minimum (the best solution) is strongly dependent on initial guess of fitted parameters, it has to be somewhat close to the final solution.

To overcome problem with initial values of parameters we have developed our own code\footnote{https://github.com/pavolgaj/OCFit} based on the using of genetic algorithms and MCMC (Markov Chain Monte Carlo) simulation. More details about our code are given % will be published 
in the upcoming paper (Gajdo\v{s} \& Parimucha~\cite{gajdos2017}. Here we will mention only brief description of main principles. The fitting of $O-C$ diagrams with our code is divided into two parts. The first part use genetic algorithms (e.g Whitley~\cite{whitley1994}) to determine initial values of fitting parameters. The second part use these values as the input to MCMC simulation (e.g. Press et al.~\cite{press2007}) which gives as a result solution with statistically significant error estimates of all parameters. 
As an input parameters, our code needs only intervals, where specific parameter can be located. User can select physically relevant intervals for each fitted parameter. With previously described approach we can find the best global solution, but its statistical significance strongly depends on number of steps in MCMC simulation, number of generations and size of population used in genetic algorithms. The discussion about selection of proper values is given in Gajdo\v{s} \& Parimucha~(\cite{gajdos2017}.

The crucial step in analysing period changes of eclipsing binaries is a setting the weights to individual observations. Minima times are determined from different types of observations, by  different instrumentation with various quality. Moreover, authors use unequal methods for minima times determinations. For our solution we choose one weight for whole group of observations obtained by one technique: visual (vis)- 1, photographic (phot) - 2, photoelectric (phe) - 10
, CCD - 10. This weighting scheme is used by many authors (e.g. Zasche et al.~\cite{zasche2009}; Liakos et al.~\cite{liakos2011}).

\begin{table*}[t]
 \caption{Parameters of the 3$^{rd}$ body orbit from $O-C$ diagram analysis of AD And and comparison with previous studies, $P$ -- orbital period of eclipsing pair, $T_0$ -- initial minimum, $Q$ -- quadratic term, $P_3$ - orbital period of the 3$^{rd}$ body, $t_{03}$ -- pericenter passage, $a\sin i_3$  -- projected semi-major axis of the orbit, $e_3$ -- eccentricity, $\omega_3$ -- the longitude of the periastron, $f(M_3)$ -- the mass function, $\chi^2$ -- sum of squares of the best fit, $\chi^2/n$ -- reduced sum of squares ($n$ -- number of data points), errors are given in parenthesis.}
\label{tab:resultsAD}
 \begin{center}
  \begin{tabular}{lccc}
    \hline\noalign{\smallskip}
	Solution			& this paper				& 	(1)					&  (2)	\\
  \hline\noalign{\smallskip} 
 $P$ [days]				&   0.98619356(6)  			& 0.98619240(14)		& 0.9861924(4)	\\ 
 $T_0$ [HJD]			& 2439002.9350(9)  			& 2439002.5733(15)		& 2439002.458(6)	\\
 $Q$   [days]			& 1.67(1.49)$\times 10^{-12}$& --					& -- 	\\
   \hline\noalign{\smallskip} 
 $P_3$ [days]   		&   4418(16) 				&  	5249				& 5220(37)\\
 $t_{03}$ [HJD]			& 2442236(537)   			& 2438813(414) 			& 2447012(175)	\\
$a\sin i_3$ [AU]		&  3.13(7) 	 				&  3.24(12)				& --\\
 $e_3$     				&  0.15(5)					& 0.30(24)		 		& 0.17(5)\\
$\omega_3$ [\degr] 		&   284(43)					& 270(50)			 	& 25(11)\\
\hline\noalign{\smallskip}
$f(M_3)$  [M$_\odot$]	&  0.209(14)	  			&  0.160(20)			& 0.183(1)\\
\hline\noalign{\smallskip}
$\chi^2$ 				&  290.596					& 	--		 			& --\\
$\chi^2/n$				&  0.723					& 	--					& --\\
\hline\noalign{\smallskip}
\multicolumn{4}{l}{(1) - Liao \& Qian~(\cite{liao2009}), (2) - Liakos et al.~(\cite{liakos2012})} 
\end{tabular}
\end{center}
\end{table*}

   \begin{figure}
   \centering
   \includegraphics[width=\columnwidth, angle=0]{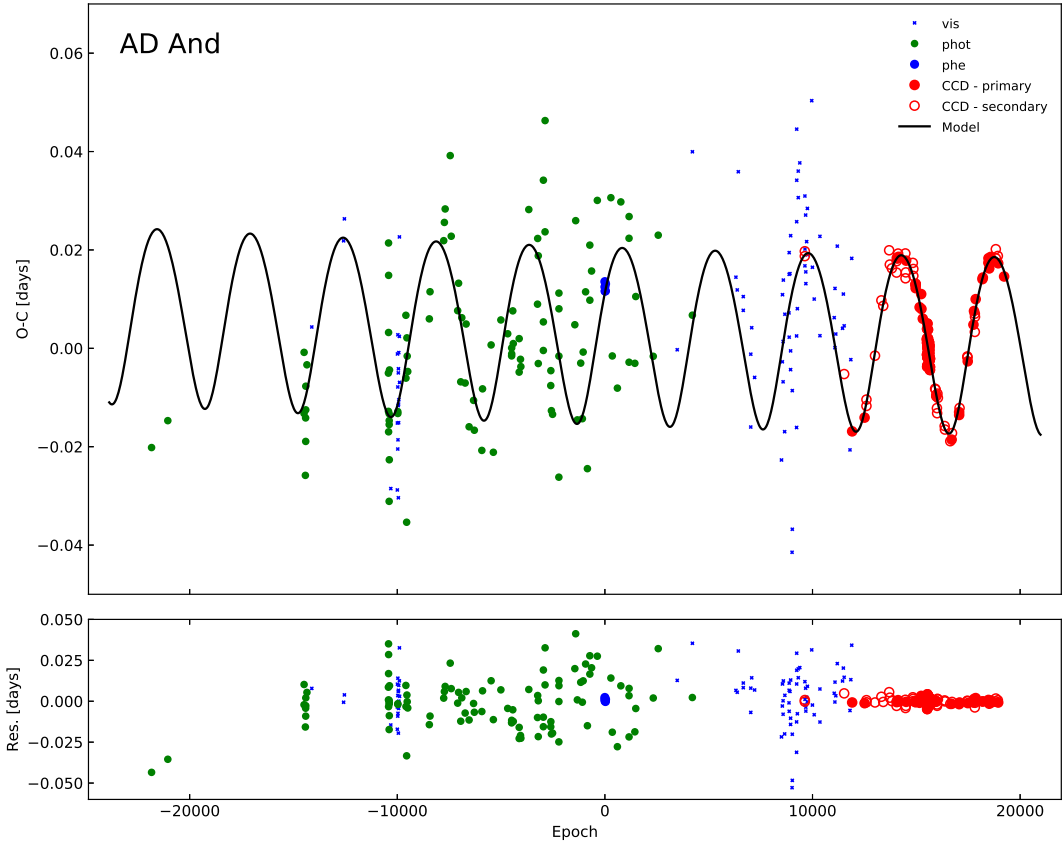}
   \caption{The $O-C$ diagram of AD~And fitted by light-time effect (upper) and the residuals after the subtraction of the the best fit (lower). Different types of observation are depicted by different points and colours. The best fit (solid black line) corresponds to values given in Tab.\ref{tab:resultsAD} }
   \label{fig:01}
   \end{figure}

\section{Results and discussion}
\subsection{AD And}
In Table~\ref{tab:resultsAD} we list our results from fitting of $O-C$ diagram for AD~And together with early published results. Our best fit solution is shown in Fig.~\ref{fig:01} 
Parameters of our solution are almost the same as in previous papers, except for the orbital period $P_3$ of the third body. It is about 2 years shorter (12.1~y, in contrast to 14.3~y) than in other solutions. This difference can be explained by the fact, that we have used much longer time interval for an analysis. Liao \& Qian~(\cite{liao2009}) and Liakos et al.~(\cite{liakos2012}) used only minima times from photoelectric and CCD observations and neglected all $O-C$ points obtained before 1990 because of their poorer quality. We used also these older photographic and visual observations even with smaller weight. Moreover, the last CCD minima times cover all circle of $O-C$ variations (see Fig.~\ref{fig:01}). 

We also detected secular period change (parameter $Q$) not mentioned by other authors. The period change corresponds to the increase of the period $dP/dt= 1.06(94) \times 10^{-4}$~sec/year and should be connected with mass transfer from the secondary component to the primary one and/or with Applegate effect (Applegate~\cite{applegate1992}). But this is not in agreement with detached configuration of the system Liakos et al.~(\cite{liakos2012}). It is necessary to note that the relative statistical error of $Q$ is almost 90\%, what degrades its significance.  
We have tried also solution with no $Q$ and we surprisingly obtained results with worse statistical significance BIC.  We cannot confirm or disprove the secular period changes and only future observation can bring light into solution of this problem.  

Our solution imply minimal $3^{rd}$ body mass (for $i_3 = 90\degr$) of $\sim 2.33$~M$_\odot$, using absolute parameters from Liakos et al.~(\cite{liakos2012}). This would yield the third light about 15\% to the total luminosity of the system. But the third light resulting from light curve analysis (Liakos et al.~\cite{liakos2012}) is about 3\%. This difference can be explained with the assumption that the $3^{rd}$ star is actually binary system with two solar-mass components as mentioned by Liakos et al.~(\cite{liakos2012}).

\subsection{TW Cas}
\begin{figure}[t]
 \includegraphics[width=\columnwidth]{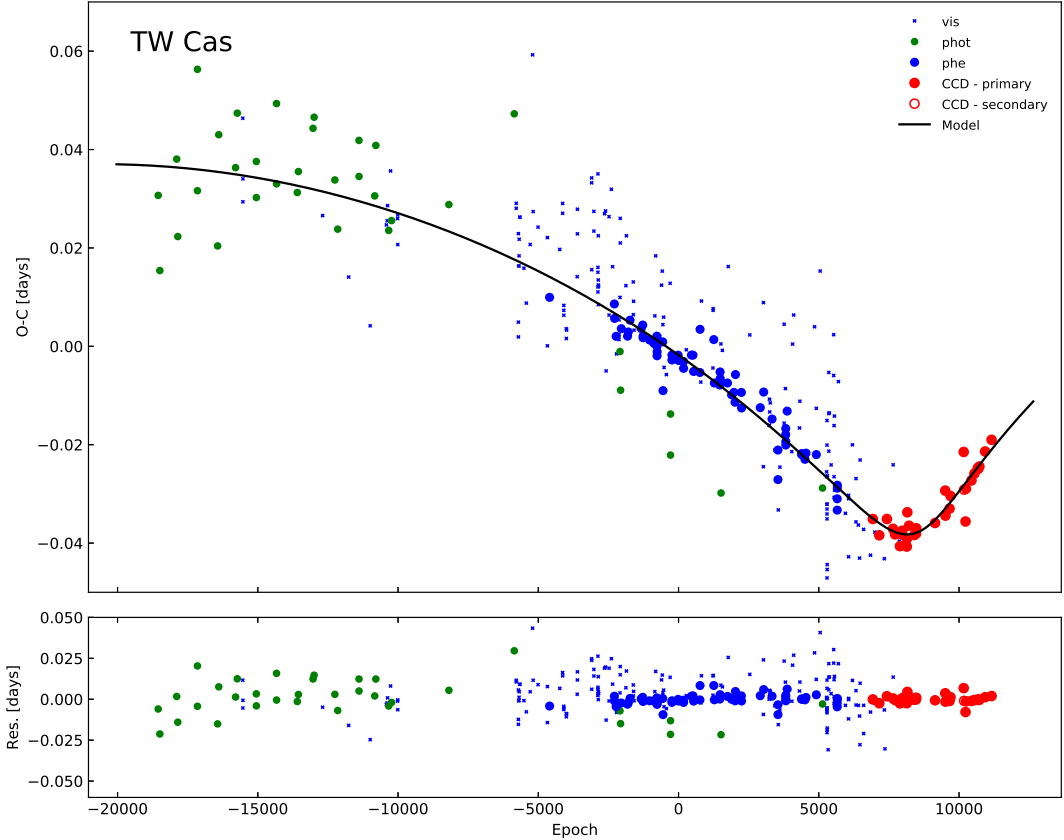}  
 \caption{The $O-C$ diagram of TW~Cas fitted by light-time effect (upper) and the residuals after the subtraction of the the best fit (lower). Different types of observation are depicted by different points and colours. The best fit (solid black line) corresponds to values given in Tab~\ref{tab:resultsTW}.}
 \label{fig:02}  
 \end{figure}

Khaliullina~(\cite{kaliulina2015}) for the first time noted, that variations of $O-C$ diagram of TW~Cas can be explained by the presence of other body in the system. This hypothesis is based on the latest CCD minima times which have different trend with respect to linear ephemeris, than older ones. Our analysis of all available minima times of this object confirmed this fact. The results from our and Khaliullina~(\cite{kaliulina2015}) studies are listed in Table~\ref{tab:resultsTW} and our best fit model is shown in Fig.~\ref{fig:02}.

We confirmed that the $3^{rd}$ body is on highly eccentric orbit ($e_3 =0.71$), although our period $P_3$ is about 3 years longer and mass function is about twice as small as previous solution. But this values are in the frame of statistical errors of these parameters. We did not reveal any secular changes caused by mass transfer and/or magnetic activity. From our model we can find minimal mass of the $3^{rd}$ body (for $i_3 = 90\degr$) to be $\sim 0.48$~M$_\odot$, using absolute parameters from Djurasevic et al.~(\cite{djurasevic2006}). If we assume that the third body is a main sequence star, its spectral type should be K6-7, absolute magnitude in $V$ passband $\sim8^m$. Its contribution to the total luminosity of the system is about 0.1\%. Photometric studies of TW~Cas did not reveal significant third light on the light curve what is in agreement with low mass $3^{rd}$ body on close to edge-on orbit.

\begin{table}[t]
 \caption{Parameters of the 3$^{rd}$ body orbit from $O-C$ diagram analysis of TW~Cas and comparison with previous analysis (for description of parameters see Table \ref{tab:resultsAD}).}
\label{tab:resultsTW}
 \begin{center}
  \begin{tabular}{lcc}
  \hline\noalign{\smallskip}
	Solution			& this paper				& 	(1)					\\
  \hline\noalign{\smallskip} 
 $P$ [days]				&   1.42832665(35)			& 1.4283273(5)			\\ 
 $T_0$ [HJD]			& 2442008.3870(15)  		& 2442008.3560(4)		\\
 $Q$   [days]			& --						& --					\\
   \hline\noalign{\smallskip} 
 $P_3$ [days]   		&   75300(2900) 			&  	74300(400)			\\
 $t_{03}$ [HJD]			& 2454255(388)   			& 2454400(200) 			\\
$a\sin i_3$ [AU]		&  6.49(68) 	 			&  7.8(1.4)			\\
 $e_3$     				&  0.71(3)					& 0.74(7)		 		\\
$\omega_3$ [\degr] 		&   284(4)					& 288(6)			 	\\
\hline\noalign{\smallskip}
$f(M_3)$  [M$_\odot$]	&  0.006(2)		  			&  0.013			\\
\hline\noalign{\smallskip}
$\chi^2$ 				&  221.043					& 	--		 			\\
$\chi^2/n$				&  0.7569					& 	--					\\
\hline\noalign{\smallskip}
\multicolumn{3}{l}{(1) - Khaliullina~(\cite{kaliulina2015})}
\end{tabular}
\end{center}
\end{table}

 \begin{table}[t]
 \caption{Parameters of the 3$^{rd}$ body orbit from $O-C$ diagram analysis of IV~Cas and comparison with previous analysis (for description of parameters see Table \ref{tab:resultsAD}).}
\label{tab:resultsIV}
 \begin{center}
  \begin{tabular}{lcc}
  \hline\noalign{\smallskip}
	Solution			& this paper				& 	(1)					\\
  \hline\noalign{\smallskip} 
 $P$ [days]				&  0.99851644(18)			& 0.99851658(12)	\\ 
 $T_0$ [HJD]			& 2440854.6186(34)  		& 2440854.6280(5) 		\\
 $Q$   [days]			& 7.33(1.69)$\times$10$^{-12}$& --					\\
   \hline\noalign{\smallskip} 
 $P_3$ [days]   		&   21700(444) 				&  	21800(500)			\\
 $t_{03}$ [HJD]			& 2439254(858) 				& 2443455(50) 			\\
$a\sin i_3$ [AU]		&  7.13(72) 	 			&  --			\\
 $e_3$     				& 0.31(10)					& 0.09		 		\\
$\omega_3$ [\degr] 		&   272(14)					& 341(3)			 	\\
\hline\noalign{\smallskip}
$f(M_3)$  [M$_\odot$]	&  0.102(3)		  			&  0.056			\\
\hline\noalign{\smallskip}
$\chi^2$ 				&  218.658					& 	--		 			\\
$\chi^2/n$				&  0.9344					& 	--					\\
\hline\noalign{\smallskip}
\multicolumn{3}{l}{(1) - Wolf et al.~(\cite{wolf2006})} 
\end{tabular}
\end{center}

\end{table}

\subsection{IV Cas}
\begin{figure}[t]
 \includegraphics[width=\columnwidth]{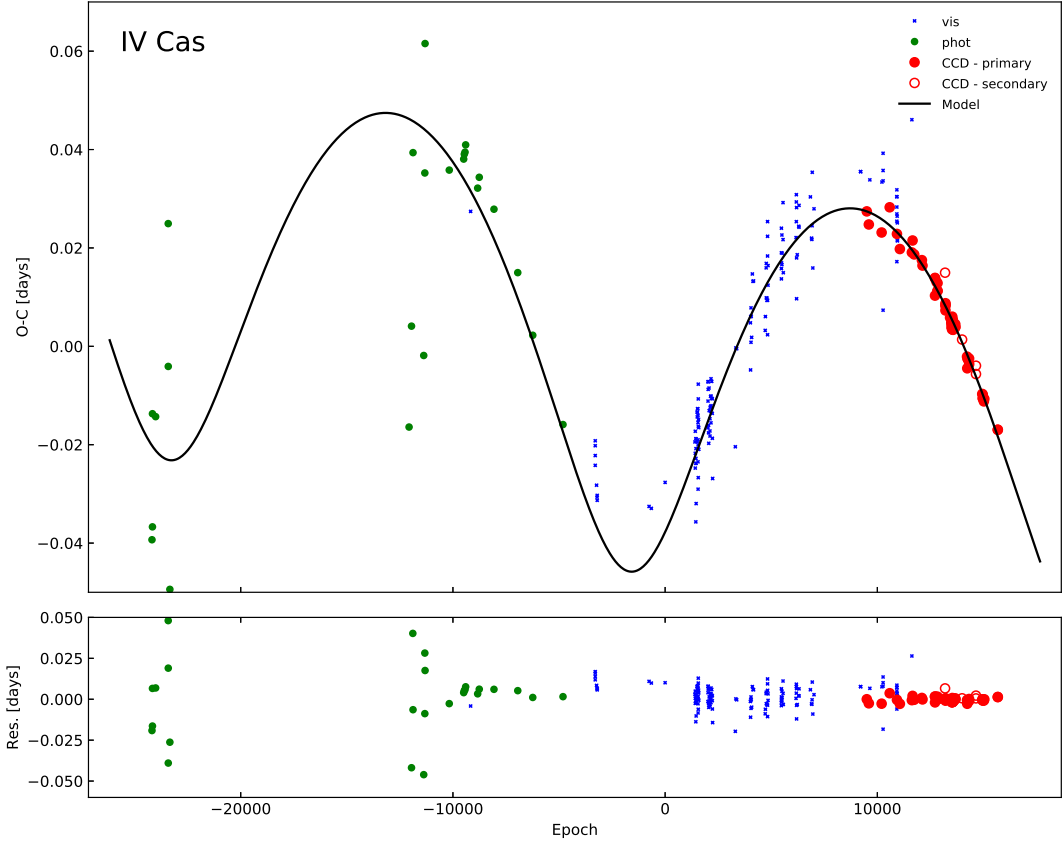}  
 \caption{The $O-C$ diagram of IV~Cas fitted by light-time effect (upper) and the residuals after the subtraction of the the best fit (lower). Different types of observation are depicted by different points and colours. The best fit (solid black line) corresponds to values given in Tab~\ref{tab:resultsIV}.}
 \label{fig:03}
 \end{figure}

Our results of $O-C$ diagram analysis of IV~Cas together with parameter's values from study of Wolf et al.~(\cite{wolf2006}) are listed in Table~\ref{tab:resultsIV} and our best solution is showed in Fig.~\ref{fig:03}. Unlike analysis of Wolf et al.~(\cite{wolf2006}) we have obtained different values for three parameters. The first one is a secular period increase $dP/dt= 4.5(6) \times 10^{-4}$~sec/year, the second one is higher eccentricity (0.31 vs. 0.09) of the $3^{rd}$ body orbit and the third one is larger mass function (0.102~M$_\odot$ vs. 0.056~M$_\odot$). The period increase could be explained by mass transfer from secondary to primary component. It is in agreement with semi-detached configuration determined by Kim et al.~(\cite{kim2010}). Significantly higher eccentricity corresponds to the shape of $O-C$ diagram with latest minima times (see Fig.~\ref{fig:03})

Our solution gives minimal mass of the $3^{rd}$ component (for $i_3 = 90\degr$) to be $\sim 1.27$~M$_\odot$, using masses from Kim et al.~(\cite{kim2010}). Assuming main sequence $3^{rd}$ body, the third light about 10\% should be observed. But Kim et al.~(\cite{kim2010}) did not report any third light  from their light-curve solution. The only realistic explanation is that the third body is actually binary star with less massive and luminous components.

\section{Conclusion}
\label{sec:conclusion}

We have analysed period variations of three Algol-type eclipsing binary stars. We used all minima times available in literature as well as newly determined from our observations and from SuperWasp archive. We used our code based on genetic algorithms and MCMC simulation. This allows us to determine fitting parameters with statistically significant errors  and also measure quality of statistical model with Bayesian information criterion.

Our new period analysis of all studied Algol-type eclipsing binaries confirmed their multicomponent nature. The third component in AD~And system is most probably also binary star with two solar-mass components, as shown by large minimal mass of this component determined from $O-C$ analysis. This is supported also by the solution of light-curve from Liakos et al.~(\cite{liakos2012}). We can speculate that orbital inclination of this binary is much lower than 90$\degr$, because we see no other set of eclipses on the light-curve. Moreover, absence of ellipsoidal variations on the light-curve caused by the second binary system suggests that this pair is detached binary on the orbit with period in the range of several days. Detected period increase is disputable and cannot be confirmed or disproved from available data. As a result, we can conclude that AD~And is a quadruple-system consisting two binaries. The first one is eclipsing pair which we observe and the second one is binary star with total mass of at least 2.33~M$_\odot$ with orbital inclination and semi-major axis that prevent us from observing other set of eclipses. 

Analysis of TW~Cas period variations approve presence of the third body in the system. This body is on highly eccentric orbit with minimal mass $\sim 0.48$~M$_\odot$, which contributes minimally to the total luminosity of the system.

Finally, $O-C$ diagram analysis of IV~Cas produced different results than previous analysis conducted by  Wolf et al.~(\cite{wolf2006}). Main discrepancies have been found in secular period increase, eccentricity and mass function. Period increase is caused by mass transfer from secondary to primary component. Light-curve solution of Kim et al.~(\cite{kim2010}) showed that secondary component fulfill its Roche lobe and this support above mentioned mass transfer. Due to excessive mass of the third body with respect to no detected third light, we can again conclude that the third body is in fact a binary system with unknown orbital parameters. Therefore, IV~Cas could be also considered a quadruple-system consisting visible semi-detached binary with pulsating primary component and the second pair composed of cool, low mass and low luminous main sequence stars of K6-7 spectral type.

\begin{acknowledgements}
This paper has been supported by the grant of the Slovak Research and Development Agency with number APVV-15-0458. This 
article was created by the realization of the project ITMS No.26220120029, based on the supporting operational Research 
and development program financed from the European Regional Development Fund. M.V. would like to thank the project VEGA 
2/0143/14.
\end{acknowledgements}

\label{lastpage}

\begin{thebibliography}{99}
%% you can type \apj for ApJ, \aap for A&A, \apss for Ap&SS, etc. Please consult
%% the macro chjaa.cls. You can also find them in aasguide.tex (AASTeX for ApJ, AJ, PASP)
%% Please follow the format of ChJAA's reference list


\bibitem[1992]{applegate1992} Applegate, J., 1992, \aj, 385, 621
% \bibitem[Brancewicz and Dworak(1980)]{brancew1980} Brancewicz, S.K. \& Dworak, T.Z.: \actaa~{\bf 30}, 501 (1980)
\bibitem[1934]{cannon1934} Cannon, A. J., 1934, Harvard College Observatory Bulletin, 897, 12
\bibitem[2006]{djurasevic2006} Djurasevic G., et al. 2006, \aap, 445, 291
\bibitem[1973]{Frieboes1973} Frieboes-Conde, H., Herczeg, T., 1973, \aaps, 12, 1
\bibitem[2018)]{gajdos2017}Gajdo\v{s}, P., Parimucha, \v{S}., 2018, submitted to A\&C
\bibitem[1981]{giuricin1981} Giuricin, G., Mardirossian, F., 1981, \aaps, 45, 499
\bibitem[1927]{gutnik1927} Guthnick, P., Preger, R., 1927, KVeBB, 1, 4
\bibitem[2001]{hilditch2001} Hilditch, R. W., 2001, An Introduction to Close Binary Stars, Cambridge University Press 
\bibitem[1975]{hill1975} Hill, G. et al., 1975, Memoirs of the Royal Astronomical Society, 79, 131
\bibitem[1952]{irwin1952} Irwin, J. B., 1952, \apj, 116, 211
\bibitem[2015]{kaliulina2015} Khaliullina, A. I., 2015, Astronomy Reports, 59, 717
\bibitem[1971]{krainer1971} Kreiner, J. M., 1971, \actaa, 21, 365
\bibitem[2004]{kreiner2004} Kreiner, J. M., 2004, \actaa, 54, 207
\bibitem[2005]{kim2005} Kim, S. L. et al. 2005, \ibvs No. 5669
\bibitem[2010]{kim2010} Kim, S. L., Lee, J. W., Lee, C. U., Youn, J. H., 2010, \pasp, 122, 1311 
% \bibitem[Kwee and van Woerden(1956)]{kwee1956} Kwee, K.K., van Woerden, H.: Bull. Astron. Inst. Neth.~{\bf 9}, 252 (1956)
\bibitem[2002]{lloyd2002} Lloyd, C., Guilbault, P., 2002, Observatory, 122, 85
\bibitem[2009]{liao2009} Liao, W., Qian, S., 2009, \na, 14, 249
\bibitem[2011]{liakos2011} Liakos, A., Zasche, P., Niarchos, P., 2011, New Astronomy, 16, 530
\bibitem[2012]{liakos2012} Liakos, A., Niarchos, P., Budding, E., 2012, \aap, 539, A129
% \bibitem[Malama(1980)]{malama1980} Mallama, A.D.: \apjs~{\bf 44}, 241 (1980)
% \bibitem[Malama(1987)]{malama1987} Mallama, A.D.: J. AAVSO~{\bf 16}, 4 (1987)
% \bibitem[Manzoori, Abbasvand and Najafinezhad(2015)]{manzori2015} Manzoori, D., Abbasvand, S., \& Najafinezhad, F.: Astron. Nachr.~{\bf 336}, 570 (2015)
\bibitem[1971]{mccook1971} McCook G. P., 1971, \apj, 76, 449
\bibitem[1940]{meshova1940} Meshkova, T. S., 1940, Variable Stars, 5, 304
%\bibitem[Mikul\'{a}\v{s}ek et \apjal.(2014)]{mikulasek2014} Mikul\'{a}\v{s}ek, Z., Chrastina, M., Li\v{s}ka, J., Zejda, M., Jan\'{i}k, J., Zhu, L.-Y., Qian, S.-B. 2014, Contrib. Astr. Obs. Sk. Pleso ~{\bf 43} 382 (2014)
\bibitem[2015]{mikulasek2015} Miku\'{a}\v{s}ek, Z., 2015, \aaps, 584, A8
\bibitem[2001]{narita2001} Narita, E., Schr\"{o}der K.P., Smith R.C., 2001, Observatory, 121, 308
\bibitem[2015]{parimucha2015} Parimucha, \v{S}., Va\v{n}ko, M. 2015, ASPC 496, 309
\bibitem[1907]{pickering1907} Pickering E. C., 1907, Astron. Nachr., 175, 91 
% \bibitem[Pikhun and Andronov(2003)]{pikhun2003} Pikhun, A.I., Andronov, I.L.:  Odessa Astr. Publ.~{\bf 16} 54
\bibitem[2006]{pollaco2006} Pollaco, D. L., et al., 2006, \pasp, 118, 1407
\bibitem[2007]{press2007}Press, W. H., Teukolsky, S. A., Vetterling, W. T., Flannery, B. P., 2007, Numerical Recipes 3rd Edition: The Art of Scientific Computing, Cambridge University Press %mesto? %tu 
\bibitem[1966]{rucinsky1966} Rucinski, S. M., 1966 \actaa~16, 307
% \bibitem[Qian(2001)]{quian2001} Qian, S.: \aj~{\bf 122}, 1561 (2001)
\bibitem[1950]{struve1950} Struve, O., 1950, \apj, 112, 184
\bibitem[2005]{sterken2005} Sterken, C., 2005, ASP Conference Series, 335, 3 
\bibitem[1940]{taylor1940} Taylor, P. H., Alexander, R. S., 1940, Publications of the Flower Astronomical Observatory, 6, 1
\bibitem[2006]{wolf2006} Wolf, M., et al., 2006, \ibvs No. 5735
%\bibitem[Wood and Forbes(1963)]{wood1963} Wood, B.D. \& Forbes, J.E.: \aj~{\bf 68}, 257 (1963)
\bibitem[1994]{whitley1994} Whitley, D., 1994, Statistics and Computing, 4, 65
\bibitem[1957]{whitney1957} Whitney, B. S., 1957, \aj, 62, 37
\bibitem[2006]{zasche2006} Zasche, P., 2006, ASP Conference Series, 349, 379
\bibitem[2009]{zasche2009} Zasche, P., et al., 2009, New Astronomy, 14, 121 
\bibitem[1913]{zinner1913} Zinner, E., 1913, Astron. Nachr., 195, 453
 
\end{thebibliography}
\end{document}